\documentclass[aps,pra,superscriptaddress,showpacs,notitlepage,onecolumn]{revtex4-1}
\usepackage{hyperref}
\usepackage{amsmath}
\usepackage{graphicx}
\usepackage{color}
\usepackage{braket}
\usepackage{verbatim}

\newcommand{\pcsadd}{Center for Theoretical Physics of Complex Systems, Institute for Basic Science, Daejeon 34126, Korea}

\newcommand{\vincaadd}{COHERENCE, Vin\v{c}a Institute of Nuclear Sciences, University of Belgrade, National Institute of the Republic of Serbia, P.O.B. 522, 11001 Belgrade, Serbia}

\newcommand{\addtuc}{School of Electrical and Computer Engineering, Technical University of Crete, Chania, Greece 73100}

\newcommand{\addcqt}{Centre for Quantum Technologies, National University of Singapore, 3 Science Drive 2, Singapore 117543}

\begin{document}

\title[Nonlinear compact localized modes in flux-dressed octagonal-diamond photonic lattice]{Nonlinear compact localized modes in flux-dressed octagonal-diamond photonic lattice}

\author{\firstname{M. G.} \surname{Stojanovi\'c}}
\affiliation{\vincaadd}

\author{\firstname{S.} \surname{G\"undo\u{g}du}}
\affiliation{\pcsadd}
\affiliation{Department of Physics, Humboldt-Universit\"{a}t zu Berlin, Newtonstrasse 15, 12489 Berlin, Germany}

\author{\firstname{D.} \surname{Leykam}}
\affiliation{\addcqt}

\author{\firstname{D. G.} \surname{Angelakis}}
\affiliation{\addcqt}
\affiliation{\addtuc}

\author{\firstname{M. } \surname{Stojanovi\'c Krasi\'c}}
\affiliation{Faculty of Technology, University of Ni\v s, 16000 Leskovac, Serbia}
\author{\firstname{M. } \surname{Stepi\'c}}
\affiliation{\vincaadd}
\author{\firstname{A.} \surname{Maluckov}}
\affiliation{\vincaadd}

\date{\today}

\begin{abstract}
	Tuning the values of artificial flux in the two-dimensional octagonal-diamond lattice drives topological phase transitions, including between singular and non-singular flatbands. We study the dynamical properties of nonlinear compact localized modes that can be continued from linear flatband modes. We show how the stability of the compact localized modes can be tuned by the nonlinearity strength or the applied artificial flux. Our model can be realized using ring resonator lattices or nonlinear waveguide arrays. 
	
\end{abstract}

\maketitle

\section{Introduction}

Flatband (FB) photonic periodic systems opened a new gate for full control of light including the manipulation of dispersion, slowing, and stopping light, which are of crucial significance in quantum technologies (cryptography, quantum computing, quantum measurements). On the other hand, the easily manageable flatband photonic systems are testbeds for exotic condensed matter phenomena associated with nearly flat bands in solid state systems~\cite{TBG,condfb,aoki}. A peculiar property of FB systems is the creation of perfectly isolated compact localized modes owing geometrically supported destructive interference \cite{flach1,flach2,flachEPL,FBluis,carlo,noncon}.

Simultaneously, switching to the world of wave vector fields and quantum systems has excited topologically oriented considerations and investigations in electronic and photonic systems \cite{Kane,Bernevig,E8967}. 
The topological photonic lattices characterized by nontrivial value of band  Chern number and bulk-edge correspondence established in the wave-vector ($k$-vectors) space challenge the researchers to explore topology-tailorable material functionalities and opportunities to go beyond those offered in the conventional materials. 

The FBs can be singular or non-singular, i.e. those with or without discontinuities in the Bloch eigenstates in the vicinity of band crossing points in the Brillouin zone (BZ), respectively~\cite{kitel,rham}. From the viewpoint of topological band theory, perfectly flat bands are 
considered as topologically trivial being characterized by a vanishing Chern number. This is related to analytic properties of the Bloch wave functions over the BZ~\cite{seidel}. Since singular FBs necessarily exhibit band crossing points their topology is ill-defined; small perturbations can open trivial or nontrivial gaps~\cite{rham}.

The most crucial property of the singular FBs is that one cannot find a complete set of compact localized states (CLSs) spanning a singular FB under periodic boundary conditions. Under open boundary conditions, the singularity is manifested by the existence of edge states, named the robust  boundary modes~\cite{rham}. These modes do not reside in a bulk band gap, but are protected by the band flatness and band crossing singularity, in contrast to topological edge states that are protected by a gap and topological charge. Thus, the FB cannot exhibit the conventional bulk-boundary correspondence \cite{rham}, which is in topological systems the correspondence between a bulk topological invariant and the in-gap boundary 
modes of the finite system.   

The two-dimensional (2D) diamond-octagon lattice (ODL) dressed by artificial flux, which can be experimentally realized by the fs-laser waveguide inscription \cite{exp1} or by tuning the networks of ring resonators \cite{exp2}, is a flexible platform for exploring FB properties \cite{BPal}. For example, an ODL model with a periodicaly-driven magnetic flux was used to study topological flatbands in Ref.~\cite{odlperiodicno}. It was shown that the flatness and topological nature of all the bands of the model can be tuned to generate topologically non-trivial flat Floquet quasi-energy bands, while their static counterparts do not support the co-existence of flatbands and nontrivial topology. 

This paper is continuation of our previous study of the flux-free nonlinear ODL, which was characterized by the band triplet consisting of two FBs and one dispersive band (DB), or four isolated DB in homogeneous or dimerized variant, respectively, in the linear limit~\cite{nashODL}. We investigated the properties of the nonlinear compact localized modes and considered the corresponding system ground state in the presence of nonlinearity.

Here, we go a step further by considering the effect of an artificial gauge field on the ODL, considering a uniform flux threading each diamond-shaped loop. This breaks the time-reversal symmetry by introducing hopping phases along the arms of each diamond plaquette~\cite{BPal}. Topological energy gaps characterized by nonzero Chern numbers emerge. As was shown in Ref.~\cite{BPal}, by fine tuning the hopping parameters topologically trivial FBs and non-trivial nearly FBs can be obtained. Such nearly flat topological bands have considered recently as a platform for various novel phenomena, including topological phase transitions induced by spin-orbit coupling and non-Abelian gauge fields \cite{41}, rich magnetic and metal-insulator phases in the presence of Hubbard interactions~\cite{42}, and quantum magnetic phase transitions induced by competition between a finite temperature and repulsive on-site interactions \cite{BPal, odllike}. 

In the paper we study the robustness of compact localized modes rising from both singular and non-singular FBs against nonlinear interactions, wondering if singularities leave their signs in the development of instability \cite{zakharov,incohmi,nlmi,extrami}. The structure of the fundamental compacton belonging to the isolated (non-singular) FBs of linear ODL is reported for the first time, according to our knowledge. We show the nearly stable nonlinear CLS propagation in the weak nonlinearity limit, which can be related to the open gaps around the FB. In addition, we study the dynamics of the linear and nonlinear CLSs when the flux is tuned away from the FB limits to test the flux-controlled manipulation of CLS. Rising from the topologically trivial FBs, the CLSs can keep their 'integrity' in the nontrivial topological environment and tune their properties via energy exchange, controllable by the applied magnetic flux. 

The article outline is as follows: In Sec. II the model of flux-dressed ODL is introduced. The focus of Section III is on the tuning of band structure by applying a magnetic flux to the diamond plaquettes. By changing the flux, topological transitions are generated. In-between the critical points the formation of singular/ non-singular topologically trivial FBs and topological nearly-flatbands is reported for selected values of the flux. The eigenmodes of FBs, compact localized modes can be extended into the nonlinear compact localized mode families in the presence of weak nonlinearity. Dynamical properties of the nonlinear CLSs in the presence of external flux changes are investigated in Section IV. The paper concludes with Section V.

\section{Model}
\label{sec:model}

\begin{figure}[ht!]
	
	\centering\includegraphics[width=12cm]{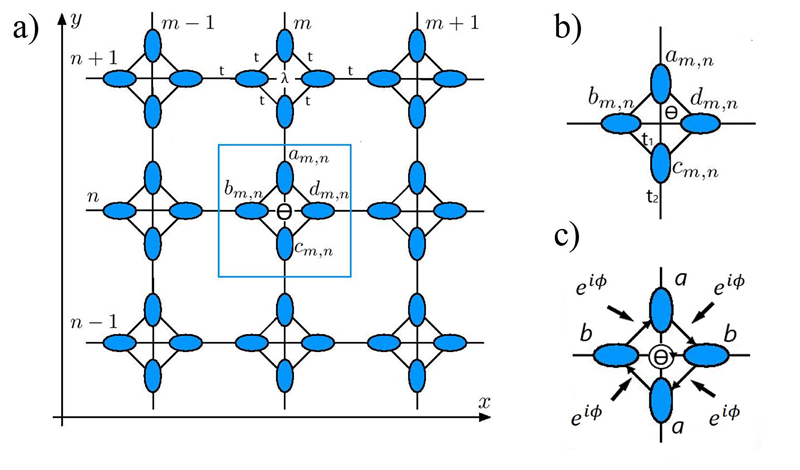}
	\caption{(a) Schematic of the ODL with artificial flux.
		The unit cell $m,n$ is bounded by a square. The hopping
		parameter along the arms of the diamond and the octagons is $t$, and the diagonal hopping is represented by $\lambda$. In the numerical simulations we select $t=\lambda=1$ for simplicity. Each diamond plaquette is threaded by a uniform external artificial flux $\Theta=4\phi$. (b) Schematic of site rings in square ring resonator lattice if Ref.~\cite{daniel} (two unit cells are presented). (c) The unit ring cell in ODL-like ring resonator lattice. The intra- and inter-site couplings are mediated by link rings with strengths $t_1$ and $t_2$, respectively. }
	\label{Model}
	
\end{figure}

The ODL geometry is schematically presented in Fig. \ref{Model}. It consists of square-diamond unit cells with all four sites linearly coupled among themselves and additionally coupled with one site from the neighboring square unit cell. The unit cells are arranged in a 2D network, so that the octagonal plaquettes alternate the squares in two orthogonal directions. We introduce the gauge field which generates the artificial magnetic field inside the squares. Flux of the artificial field modifies the couplings between unit cell sites, i.e. incorporates an Aharonov-Bohm (AB) phase factor to hopping parameter $t\rightarrow t \exp(\pm i\phi)$.  The total system Hamiltonian consists of linear $H_L$ and nonlinear $H_{NL}$ parts:
\begin{eqnarray}
H&=&H_L+H_{NL}\\
H_L({\vec k})&=&\sum_{\vec{k}}\Psi^+_{\vec{k}} H(\vec{k})\Psi_{\vec{k}}\\
H_{NL}&=& g\, \mathrm{diag}[|\psi_a(\vec{r})|^2, |\psi_b(\vec{r})|^2,|\psi_c(\vec{r})|^2, |\psi_d(\vec{r})|^2]
\end{eqnarray}
where $\Psi^+_{\vec{k}}=(\tilde{\psi}_a(\vec{k})\,\tilde{\psi}_b(\vec{k})\,\tilde{\psi}_c(\vec{k})\,\tilde{\psi}_d(\vec{k}) )^+$, $g$ is the nonlinear interaction strength, so that the intensity per unit cell is  $I(\vec{r})=\sum_{j}|\psi_j(\vec{r})|^2,\, j=a,b,c,d$. The total Hamlitonian is characterized by $H(\Theta)=H(\Theta+2m\pi)$, where $m$ is arbitrary integer, $\Theta=4\phi$ is the threaded flux, and the fundamental period is $T_{\Theta}=2\pi$.

The linear ODL  Hamiltonian can be written in matrix form:
\begin{eqnarray}
H_L=
\begin{bmatrix}
0 & te^{i \phi}  & \lambda+te^{-ik_{y}} & te^{-i \phi}\\
te^{-i \phi/4} & 0 & te^{i \phi} &  \lambda+te^{ik_{x}} \\
\lambda+te^{ik_{y}}  & te^{-i \phi} & 0 & te^{i \phi}\\
te^{i \phi} &  \lambda+te^{-ik_{x}}  & te^{-i \phi} & 0
\end{bmatrix}
\end{eqnarray} 
where $\vec{k}=(k_x,k_y)$ is the two component wavevector, $t$ and $\lambda$ are coupling coefficients. Solving the eigenvalue problem of linear Hamiltonian gives the corresponding eigenvalue bands $\beta_n(k_x,k_y)$, with band index $n=1,2,3,4$, and the corresponding eigenmodes.

The light propagation through the ODL lattice can be modeled in real space by a set of linearly coupled difference-differential Schr\"{o}dinger equations with the on-site cubic nonlinear terms:
\begin{eqnarray}
i \dot{a}_{m,n}&+&t(b_{m,n} e^{i \phi}+d_{m,n} e^{-i \phi}+c_{m,n+1})+\lambda c_{m,n}+g |a_{m,n}|^2a_{m,n}=0,\nonumber\\
i  \dot{b}_{m,n}&+&t(a_{m,n}e^{-i \phi}+c_{m,n}e^{i \phi}+d_{m-1,n})+\lambda d_{m,n}+g |b_{m,n}|^2b_{m,n}=0,\nonumber\\
i  \dot{c}_{m,n}&+&t(b_{m,n} e^{-i \phi}+d_{m,n}e^{i \phi}+a_{m,n-1})+\lambda a_{m,n}+g |c_{m,n}|^2c_{m,n}=0, \nonumber\\
i  \dot{d}_{m,n}&+&t(a_{m,n}  e^{i \phi}+c_{m,n} e^{-i \phi}+b_{m+1,n})+ \lambda b_{m,n}+g |d_{m,n}|^2d_{m,n}=0,
\label{jednacineModela}
\end{eqnarray}
where the discrete wave function $\psi_{m,n}$ is represented by components $(a\, b \, c \, d)_{m,n}^T$, which are complex mode amplitudes of the total wave field (4-component spinor), $n,m$ are the cell indices. The total number of cells is $N^2$.  In the following, for simplicity we fix $\lambda = t$ and measure energies in units of $t$ by setting $t=1$. 

The tight-binding ODL Hamiltonian can be related to experimentally achievable platform of translation invariant optical resonator network in analogy with the 2D photonic system presented in Ref.~\cite{daniel}. We can consider the unit cell of resonators consisting of resonant 'site rings' coupled via off-resonant 'link-rings' positioned to mimic the intra- ($t_1$) and inter-cell hopping ($t_2$) shown in Fig.~\ref{Model}(b). 
The corresponding model can be obtained following the procedure considered in detail in Ref.~\cite{daniel}. Briefly, the threaded flux $\Theta$ is controlled by the detuning of the off-resonant link rings. Coupling mediated by off-resonant links is described by the effective coupling strength $t_1=t\, \mathrm{csc}(\Theta/2)$ where $t$ is the coupling strength for a perfectly anti-resonant link $\Theta=\pi$. All sites then have an additional on-site potential of $t\, \mathrm{cot}(\Theta/2)$, where for an infinite lattice this potential can be removed by a simple shift of the energies and is thus irrelevant. 
The inter-cell couplings $t_2$ can be implemented either directly or by anti-resonant links when it has a form $t_2=J \mathrm{csc}(\Theta/2)$ with $J$, in general, another coupling strength. In our case $J=t$, and diagonal coupling inside the unit cell is identical to $t_1$. After rescaling the energies with respect to the common factor, $\mathrm{csc}(\Theta/2)$, the linear part of ODL tight binding model is reproduced in the form of Eq.~\eqref{jednacineModela} with $t=\lambda$. 
The only exception is $\Theta=0$, where the link rings are also resonant. To implement the flux-free limit, one can alternatively use arrays of weakly-coupled optical waveguides, where the cross-coupling is mediated by a detuned waveguide~\cite{daniel, coupledmode,coupledmode2,coupledmode3,coupledmode4}.

Concluding this consideration of the ODL risen on the ring-resonator platform \cite{exp2,daniel} we point out the advantage of consisting in the huge manageability with respect to the light intensity and robustness to losses. However, the weak point can be the requirement for a densely-packed network of resonators in the ODL structure, which may be challenging to implement. 

Finally, we would like to mention the possibility for the experimental realization of gauge field effects in waveguide-based ODLs~\cite{exp1,hafezi,expp}. The problem generically related to this approach is that the high light intensities and related strong nonlinear effects can cause high losses, whose prevention is an unresolved problem. On the other hand, very recent results inducing the dipolar waveguides in the diamond plaquettes to facilitate the AB phase can be promising route to prevent the problem with auxiliary waveguides~\cite{rodrigo}. 

\section{Tuning the band structure in linear lattice by flux}

\begin{figure}[ht]
	\centering\includegraphics[width=12cm]{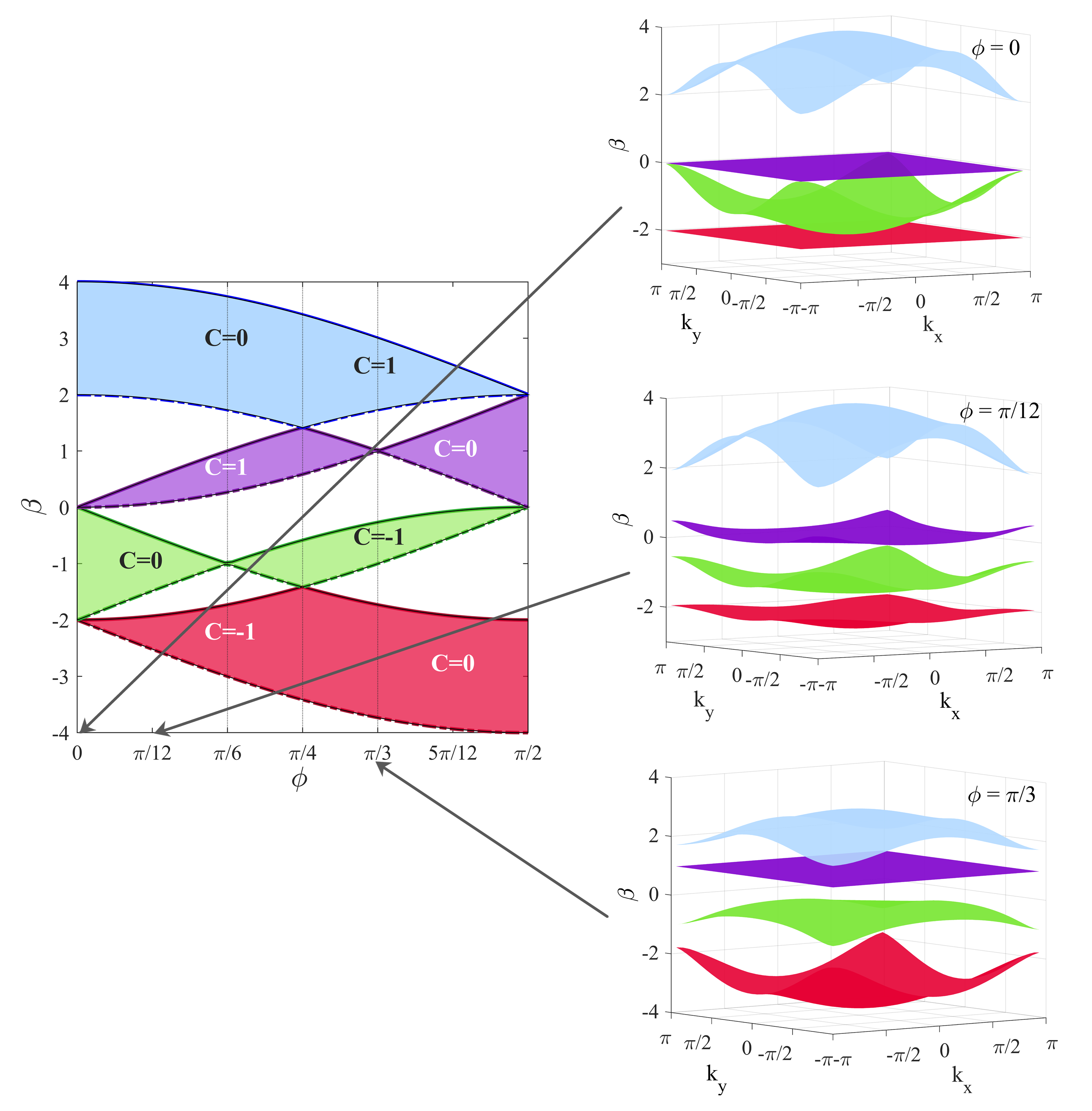}
	\caption{Energy bands $\beta$ of the flux-dressed ODL as a function of the hopping phase $\phi$ and corresponding values of the Chern number. For each $\phi$ value the lowest and highest energy $\beta(\vec{k})$ is plotted for four bands, which are denoted by different colors, within the half of the fundamental period $T_{\phi}=\pi$. The topological transition point at $\phi=\pi/4$ and flux values where perfect gapped FBs occur ($\phi=\pi/6, \beta=-1$ and $\phi=\pi/3, \beta=1$) are shown by vertical dotted lines. The bands' profiles are presented on the right hand side to illustrate the appearance of topologically trivial perfect FBs and topologically nontrivial nearly FBs.}
	\label{fbs}
\end{figure}

First we consider the linear band structure of the ODL. The linear eigenstates are Bloch functions satisfying $H_L u_n(\vec{k}) = \beta u_n(\vec{k})$, where $\beta$ is eigenenergy. Using the flux $\phi \in [0, \pi]$ as a tuning parameter the flux-dressed ODL provides a flexible platform for tuning transitions from topologically trivial to nontrivial bands, as shown in Fig.~\ref{fbs}.

To compute the topological band invariants we follow the approach of Ref.~\cite{fukui} based on the computation of Bloch functions and corresponding link variables in any gauge at discrete points within the first BZ. Here we start from the $4$x$4$ spinor wave function of the $n$th band and form the link variable
\begin{equation}
U_{n}({\vec k}_l)=\frac{1}{N_{n}({\vec k}_l)} <u_n({\vec k}_l)|u_n({\vec k}_l+\vec{\mu})>,
\end{equation}
where $N_{n}(k_l)=|<u_n({\vec k}_l)|u_n({\vec k}_l+\vec{\mu})>|$ is a normalization constant, $\vec{k}_l$ denotes the discretized grid of sampled lattice points, and $\vec{\mu} = \vec{1}, \vec{2}$ denotes a displacement in the $\vec{k}$-space lattice. The link variable denotes a complex phase factor of unit modulus which is well defined over the whole first BZ except at singular points ($N_{n}(k_l)=0$), which can be avoided by an infinitesimal shift of the grid of sampled $\vec{k}$ points. In other words, the gap-opening condition is necessary to adopt the previous procedure and define unique band Chern number. The next step is to calculate the field strength 
\begin{equation}
F_{12}(\vec{k}_l)=\ln({ U_1(\vec{k}_l)U_2(\vec{k}_l+\vec{1})U_1(\vec{k}_l+\vec{2})^{-1}U_2(\vec{k}_l)^{-1}}),\quad -\pi<\frac{1}{i}F_{12}(\vec{k}_l)\le \pi.
\end{equation} 
and obtain the Chern number of the band of interest as
\begin{equation}
C_n=\frac{1}{2\pi i}\sum_l\, F_{12}(\vec{k}_l).
\end{equation}
It is proven that the $C_n$ is strictly an integer for arbitrary lattice spacing \cite{fukui}. 

We are here particularly focused on the FBs, so to characterize the topological band flatness we introduce the flatness parameter \cite{flat}, defined as the ratio between the minimum band gap ($\Delta)$ and bandwidth $(W)$: $F=\Delta/W$.

In the absence of flux ($\phi=0)$ all bands are topologically trivial.  As illustrated in Fig. \ref{fbs}, two of them are fully degenerate (dispersionless) FBs with energy: $\beta=0$ and $-2$. The first one touches the embedded DB at the boundaries of the BZ, and the second one touches the same DB at the center of BZ~\cite{nashODL}. Corresponding band crossings are conical and parabolic, respectively, similar to the Lieb \cite{liebikagome,lieb} and Kagome lattices \cite{liebikagome}. Therefore both FBs are singular with singularities in the crossing points with the DB. The fourth band is dispersive and gapped from the FB-DB-FB triplet.

The AB modulation of hopping can be associated with the topological phase transition after band gap opening at critical flux values, as a consequence of the time-reversal symmetry breaking. The openinig of gaps between the FBs and DB in the corresponding fluxless triplet by increasing the value of flux is sketched in Fig.~\ref{fbs}. This is accompanied by the formation of topological edge modes and nontrivial topological bands with respectively $C=-1,\, 0,\, 1,\,0$.

\begin{figure}[ht]
\centering	\includegraphics[width=6cm]{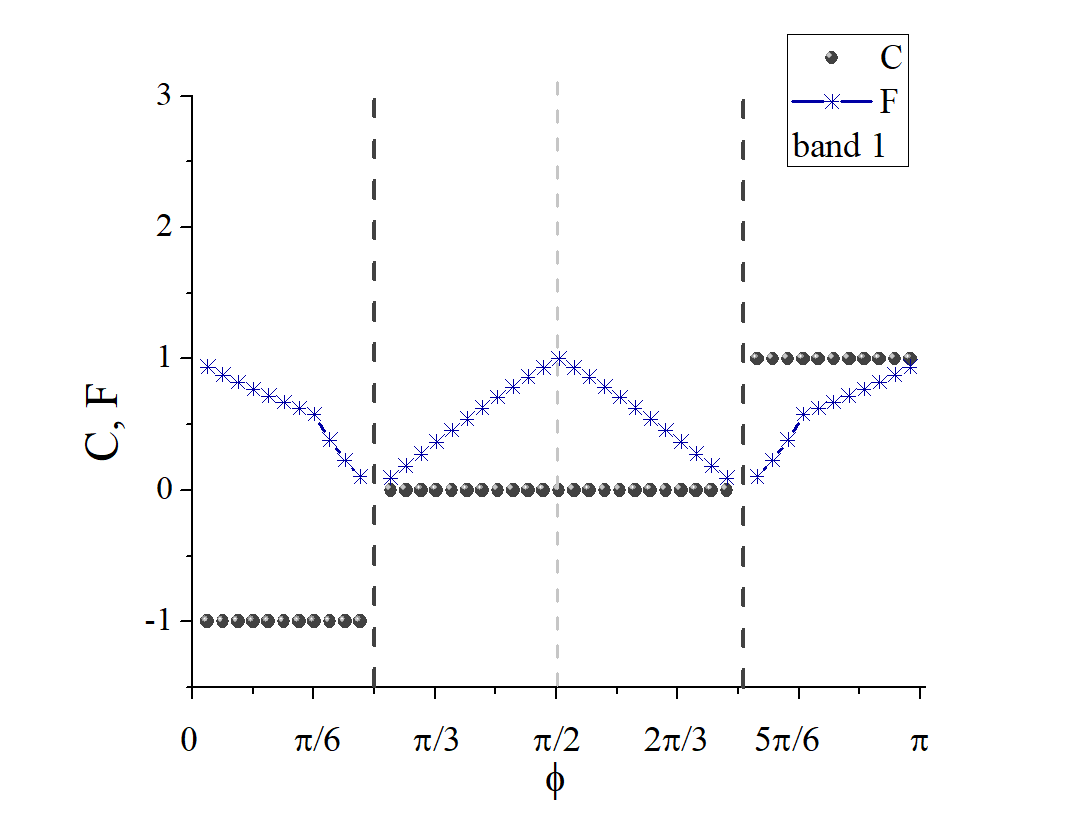}
	\includegraphics[width=6cm]{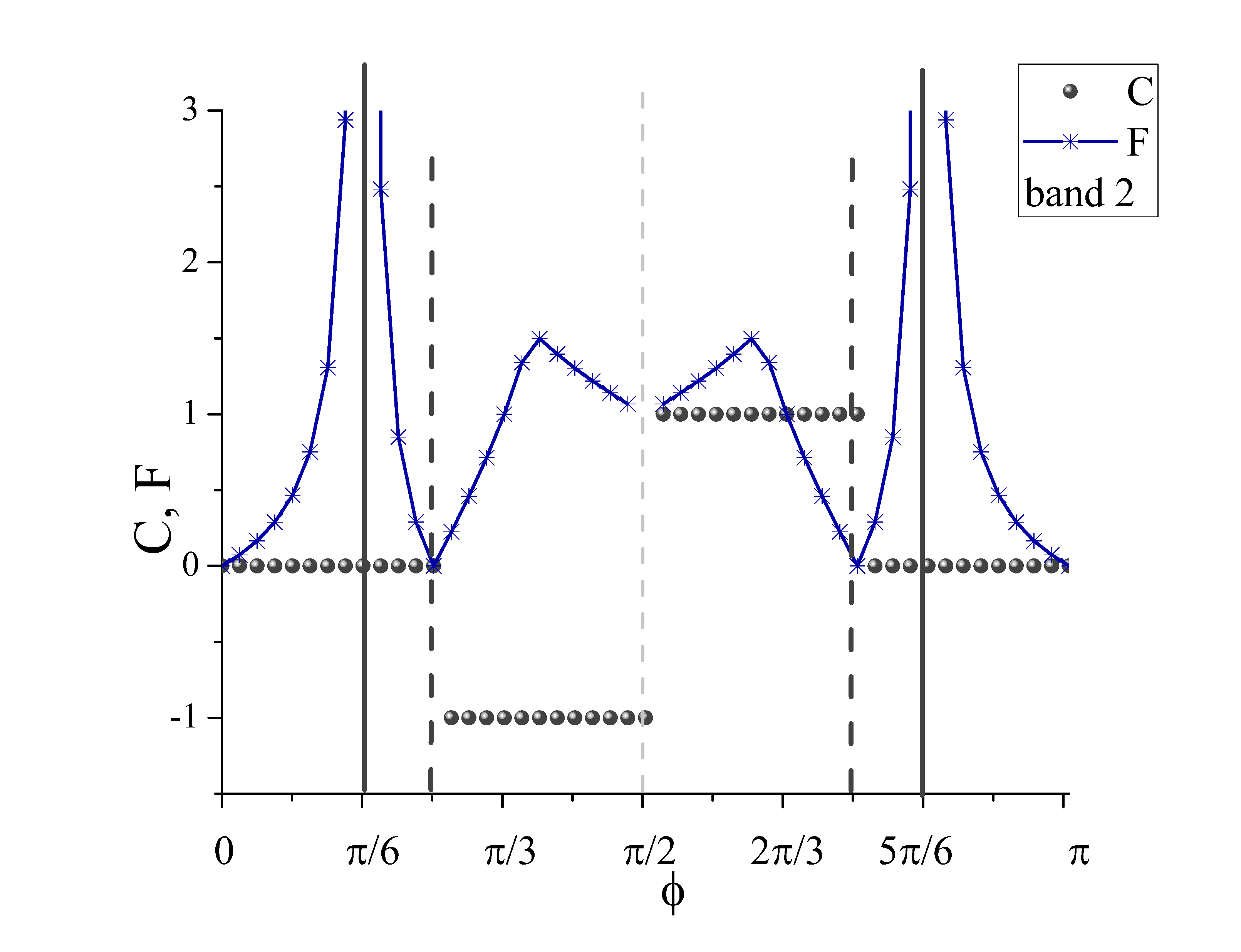}
\centering	\includegraphics[width=6cm]{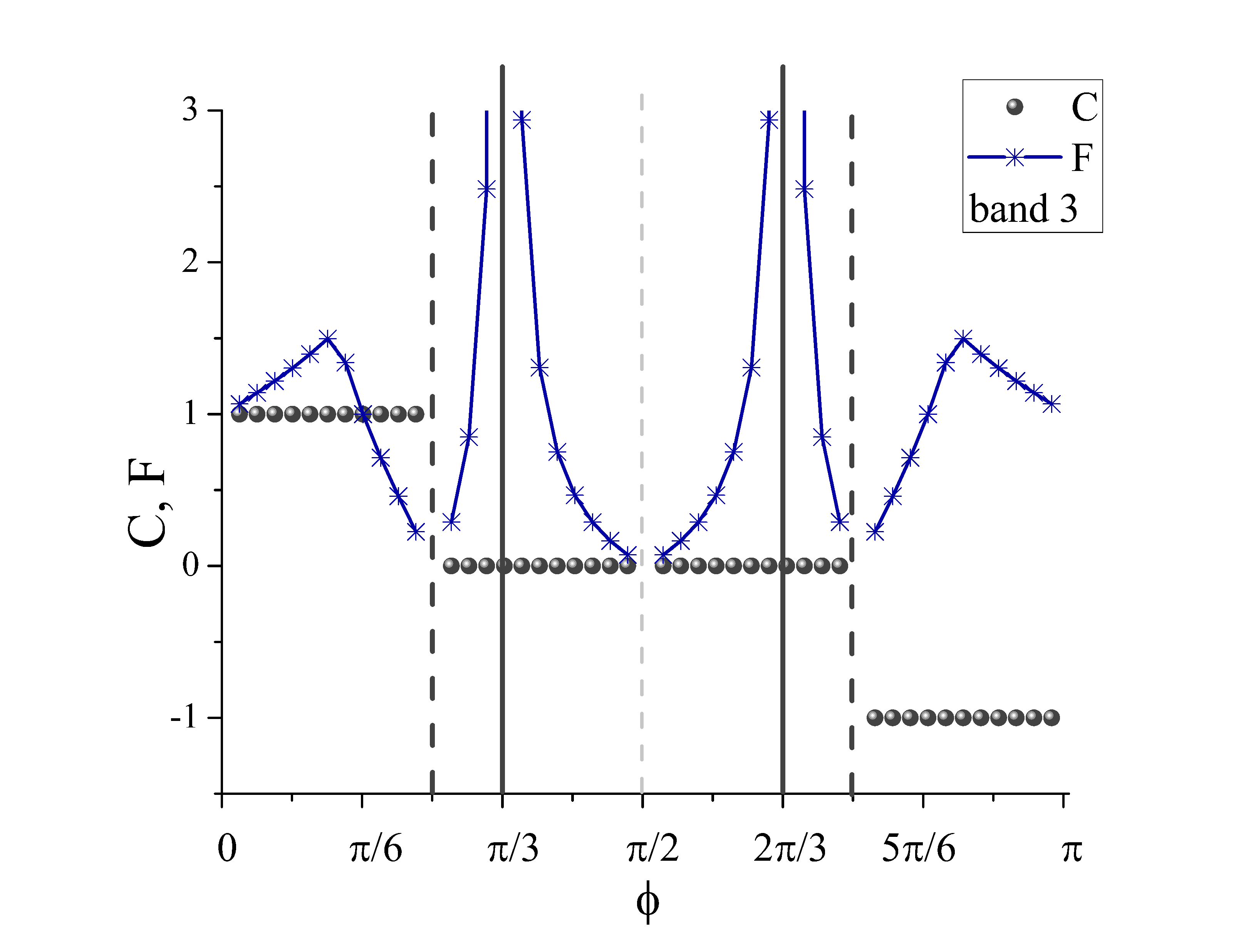}
	\includegraphics[width=6cm]{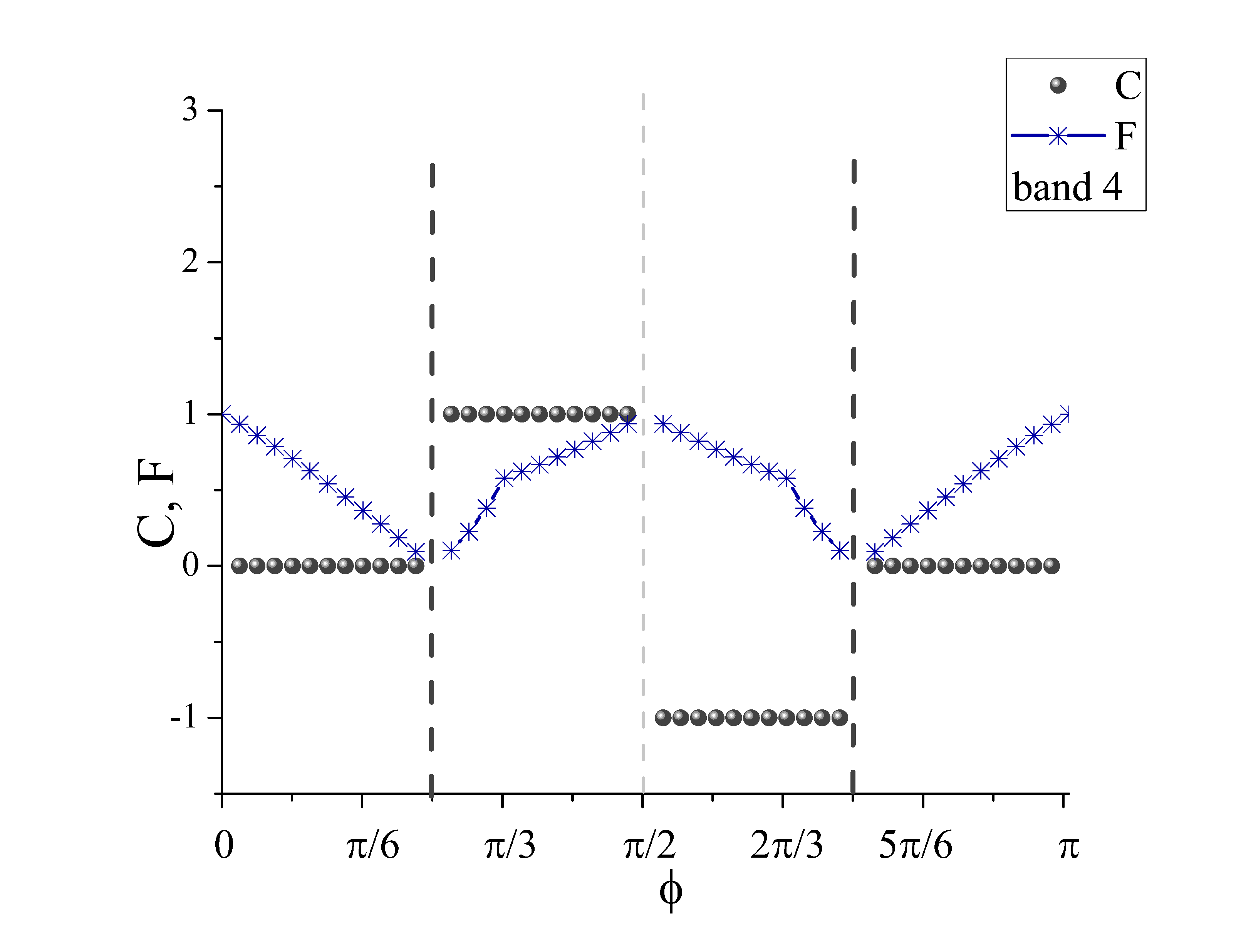}
	\caption{Values of bands flatness ($F$) and Chern number ($C$) vs. flux. Transition points correspond to $\phi=0,\, \pi/4,\, \pi/2,\,3\pi/4$ (dashed vertical lines), while FBs are created at $\phi=0,\,\pi/2,\, \pi$ (singular FBs), $\phi=\pi/6\, \pi/3$ and $\phi=2\pi/3$ (isolated FBs- solid vertical lines). Topologically nontrivial nearly FBs can be found in the area between $\phi=0$ and $\phi=\pi/6$.}
	\label{chernflat}
	
\end{figure}

The width of bands and the band gaps can be estimated from the curves in Fig.~\ref{fbs} at each $\phi$, giving the values of band flatness which are presented in Fig.~\ref{chernflat}. Non-singular perfect FBs are characterized by $F \gg 1$, while the flatness of singular FBs is not uniquely defined being the ratio of two vanishing numbers: bands 1 and 3 at $\phi=0$, and bands 2 and 4 at $\phi=\pi/2$, etc.  On the other hand, the vanishing $F$ at finite $W$ denotes the band crossing, and can help to determine the Dirac points and the topological phase transition points. In Fig.~\ref{chernflat} the last case corresponds to $\phi=\pi/4$ and $3\pi/4$, as denoted by the vertical dashed lines. 

By increasing the value of flux in Fig.~\ref{fbs} we observed band spectra with triplet of isolated nearly flatbands (for example, $\phi\approx \pi/12$) characterized  by finite $F$ and the Chern numbers arranged as $-1,0,1$. The fourth band is a topologically trivial DB. At $\phi=\pi/6$ the band characterized by $C=0$ transforms into the nonsingular fully degenerated-perfect FB at $\beta=-1$. By further change of flux, bands form two doublets connected at the edges of the BZ: $\phi=\pi/4$ in Fig.~\ref{fbs}. New gap openings appear with increasing flux and the previous scenario is repeated. When $\phi = \pi/3$ there is one non-singular perfect trivial FB at $\beta=1$ and three DBs with $C=0,-1,1$ as can be seen in Fig. \ref{fbs}. At $\phi = 5\pi/12$ there are three nearly FBs with $C=-1,0,1$ and one trivial DB. Finally, at $\phi=\pi/2$ there is a FB-DB-FB triplet plus isolated DB are formed as a mirror symmetric structure to the fluxless band structure, $C=(0,1,0,-1)$. The band spectrum pattern repeats with the periodicity $T_{\phi}=\pi$. The whole band structure cycle is related to the AB flux modification of hopping inside the diamond plaquettes surrounding the octagon rings in the ODL.

\section{Compact localized modes}

The perfect, fully degenerate FBs can host the compact localized eigenmodes - compactons - which are created by the destructive interference of superpositions of extended (linear) Bloch modes, which are the eigensolutions of the linear ODL potential. Compactons do not posses tails, being strictly isolated within the part of lattice and do not interact with the background. In the flux-dressed ODL the perfect FBs are formed at $\phi_{FB}=(m-1)\pi/6,\, m \in [1,7]$ within the fundamental period. The FBs at $0,\,\pi/2,\pi$ are singular, while those in-between are gapped. 

\begin{figure}[ht!]
\centering	\includegraphics[width=10cm]{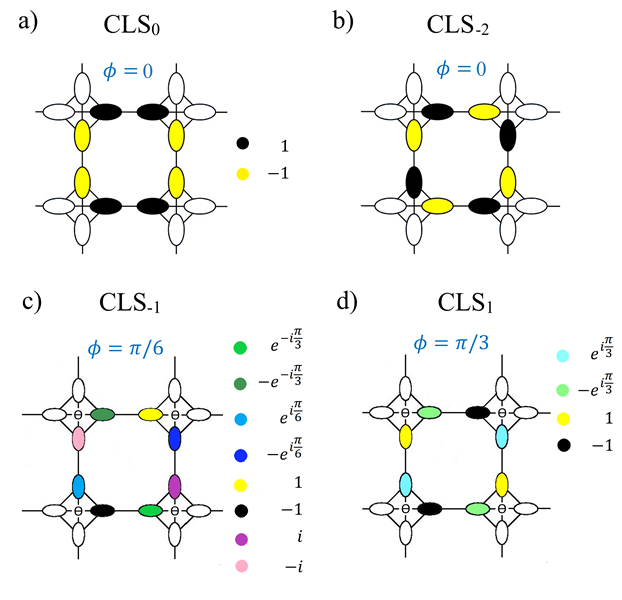}
	\caption{Fundamental compactons hosted by the octagonal plaquette: (a) $\phi=0,\, \beta=0$ ($CLS_{0}$), (b) $\phi=0,\, \beta=-2$ ($CLS_{-2})$, (c) $\phi=\pi/6,\, \beta=-1$ ($CLS_{-1})$, and (d) $\phi=\pi/3,\, \beta=1$ ($CLS_{1})$. Total flux $\Theta=4\phi$.}
	\label{compacton}
\end{figure}

In Fig.~\ref{compacton} the fundamental compactons of the singular and non-singular FBs are shown together. Compactons are of $U=4$ class, i.e. they are shared by four neighboring diamond-like unit cells. Nonzero amplitudes are realized over the octagon ring. Four compacton structures are linearly independent and can be interpreted as different realizations of an entity characterized by peculiar amplitude-phase redistribution between the eight sites, determined by the flux modulated hopping in the four diamond plaquettes at octagon corners. The AB caging appears at certain values of the AB flux fulfilling the selection rule $\phi=\phi_{FB}$. At these flux values the AB caging provides the destructive interference conditions ensuring the octagonal compact ring to be restored. 

Each linear combination of fundamental compactons for certain flux value is also eigensolution of the corresponding FB. In other words, the sets of $N^2$ fundamental compactons distributed over successive unit cells through the lattice form compact but not orthogonal FB basis in non-singular case, while in the singular one the CLS set have to be extended by the set of non-contractible loop states in order to construct the FB compact basis~\cite{rham}. 

The addition of local, on-site, power-law nonlinearity does not affect the conditions for geometry provided destructive interference \cite{carlo} or the CLSs' existence. However, for certain ranges of the system parameters, nonlinearity can give rise to instability, resulting of transfer of energy from the CLS to Bloch waves~\cite{zakharov}. The mixing efficiency is directly related to the band flatness, rate of degeneracy, band gap width, presence of the singular (band crossing) points, strength and type of the nonlinearity. 

Our aim here is to consider the effect of the nonlinearity-induced mixing on the CLS  persistence by observing and qualifying the CLS dynamical response on the development of nonlinearity-induced instability. On the one hand, this response can be used to measure the properties of the underlying band, e.g. by detecting signatures of band reconnections and topological phase transitions. On the other hand, parameter regimes in which the CLS are stable may allow them to be used as reliable information carriers in photonic lattice-based devices. 

To quantify the mode stability we follow the evolution of the participation number $P$, defined as
\begin{equation}
P=\frac{{\cal{P}}^2}{\sum_{m,n} ({|a_{m,n}|}^4+{|b_{m,n}|}^4+{|c_{m,n}|}^4+{|d_{m,n}|}^4)},
\label{eq:PN}
\end{equation}
where $\cal{P}$ is the total power (norm). 
For a compact mode with homogeneous amplitude $P$ is proportional to the number of
sites on which the mode is localized ($P\approx 8$).

Additional information about the CLS persistence is provided by calculation of the normalized magnitude of field overlap $\rho$:
\begin{equation}
\rho(z)=|<\psi(z)|\psi(0)>|^2
\label{imb}
\end{equation}
where $|\psi(z)>$ and  $|\psi(0)>$ denote the $4N^2$ dimensional field's vector whose components are mode amplitudes over the lattice sites at arbitrary $z$ and $z=0$, respectively. Discrepancy of the field overlap from unity signs the level of the CLS interaction with the environment and can be related with the loss of its stability.

To characterize the initial phase of the instability development of the CLSs, we applied the linear stability analysis (LSA). The details of the LSA procedure are shown in the number of papers (see e.g. Ref.~\cite{nashneki}). Briefly, initially slightly perturbed CLS mode is injected into the system:
\begin{equation}
\Psi_{m,n}(z)=(\psi_{m,n}(0)+\delta_{m,n}(z))\exp{(-i\beta_{NL} z)},
\label{lsa1}
\end{equation}
where $\psi_{m,n} (0)$ is the corresponding CLS with eigenenergy $\beta_{NL}$, and $\delta_{m,n}$ is small/slow complex perturbation. The nonlinear equation of small/slow perturbations is obtained after the substitution of Eq.~\eqref{lsa1} into the system of equations~\eqref{Model} and linearizing with respect to the smallness of perturbation ($|\delta|/|\Psi_0|<<1$). The resulting linear evolution equation of perturbations is then solved as the quasi-eigenvalue problem of perturbation. 
The corresponding eigenvalues determine the perturbation behavior and mode stability. Two instability types can be distinguished - the exponential and oscillatory one. The highest (absolute) value  of the real part of the complex eigenvalue  (in general),  determines the mode instability growth rate. Regarding the analysis of the nonlinear localized modes formation in photonic media, the LSA is provided to clarify the stability properties of the nonlinear Bloch modes/ plane waves which are unstructured beams. Here the focus is on the stability of nonlinear CLSs, which are structured light beams originating from the linear, fully degenerated bands. This and the fact that the corresponding linear CLS basis is compact only in the case of isolated FBs ~\cite{rham}, and consists of non-orthogonal elements demand caution in the application of the perturbative LSA. 

\subsection{Dynamical properties of nonlinear compact localized mode}

We can reinterpret our setup - the linear flux-dressed ODL - as the flux driven photonic system whose band topology successively goes through the phase characterized by the perfect FBs (zero Chern number), the topologically nontrivial nearly FBs, the gapped prefect FBs (zero Chern number), and so on. The full driving cycle is closed after $\Theta=2\pi$ when the band structure is the mirror image of the initial flux-free band structure. The driving formally introduces the additional (phase related) degree of freedom. The CLSs - $8$ site-octagonal rings are 'observables' in our approach. Any change in the band structure driven by flux imprints into the CLS intra-site phase distribution. At particular flux values fulfilling 'the selection rule' for the AB hopping, i.e. $\phi$ equal the integer multipliers of a single 'flux quantum' $\pi/6$ the compacton with peculiar phase distribution is formed, as was shown in Fig.~\ref{compacton}. Formally we may claim that the $8$ ring compactons are different phase states of the same object induced by the driving flux.  

Technically our approach is based on the propagation of the selected CLS structure through the nonlinear lattice. We numerically, by utilizing the Runge-Kutta procedure of the $6$th order, simulate the CLS mode propagation through the finite lattice with periodic and open boundary conditions regarding the problem we are addressing. The unavoidable noise in the realistic waveguide networks are numerically modeled by adding initially the uniform random perturbations on the mode amplitude and phase \cite{nashneki}. The results presented in the following are obtained for maximum value of the random perturbation field amplitude of the order $0.01$. To clarify the mode dynamics, and indirectly the properties of the band spectrum of lattice, we present the participation number, and the magnitude of the state overlap. Due to completeness, we mention the results of the LSA applied in the framework of the nonlinear CLS. 

\begin{figure}[ht!]
\centering	\includegraphics[width=7cm]{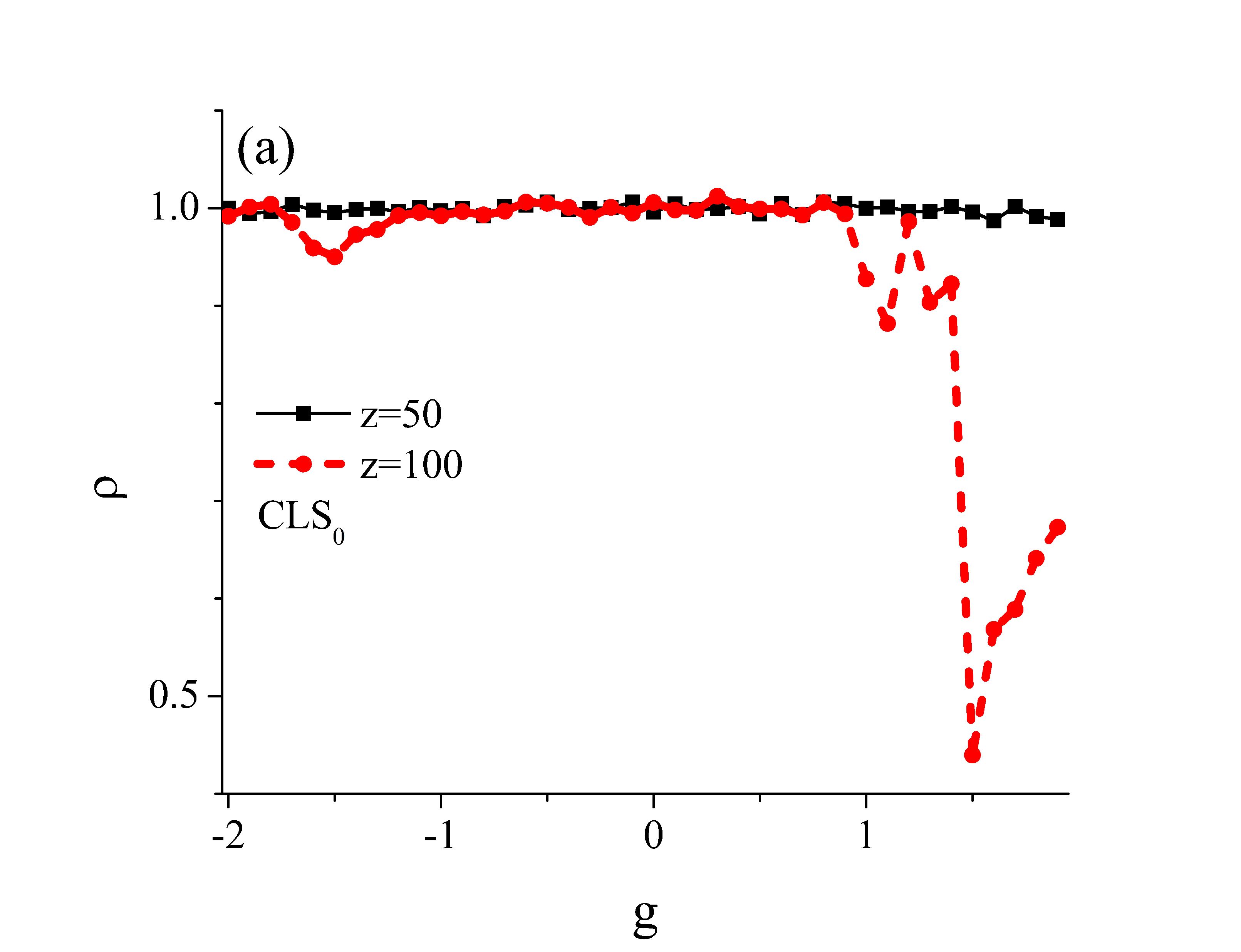}
	\includegraphics[width=7cm]{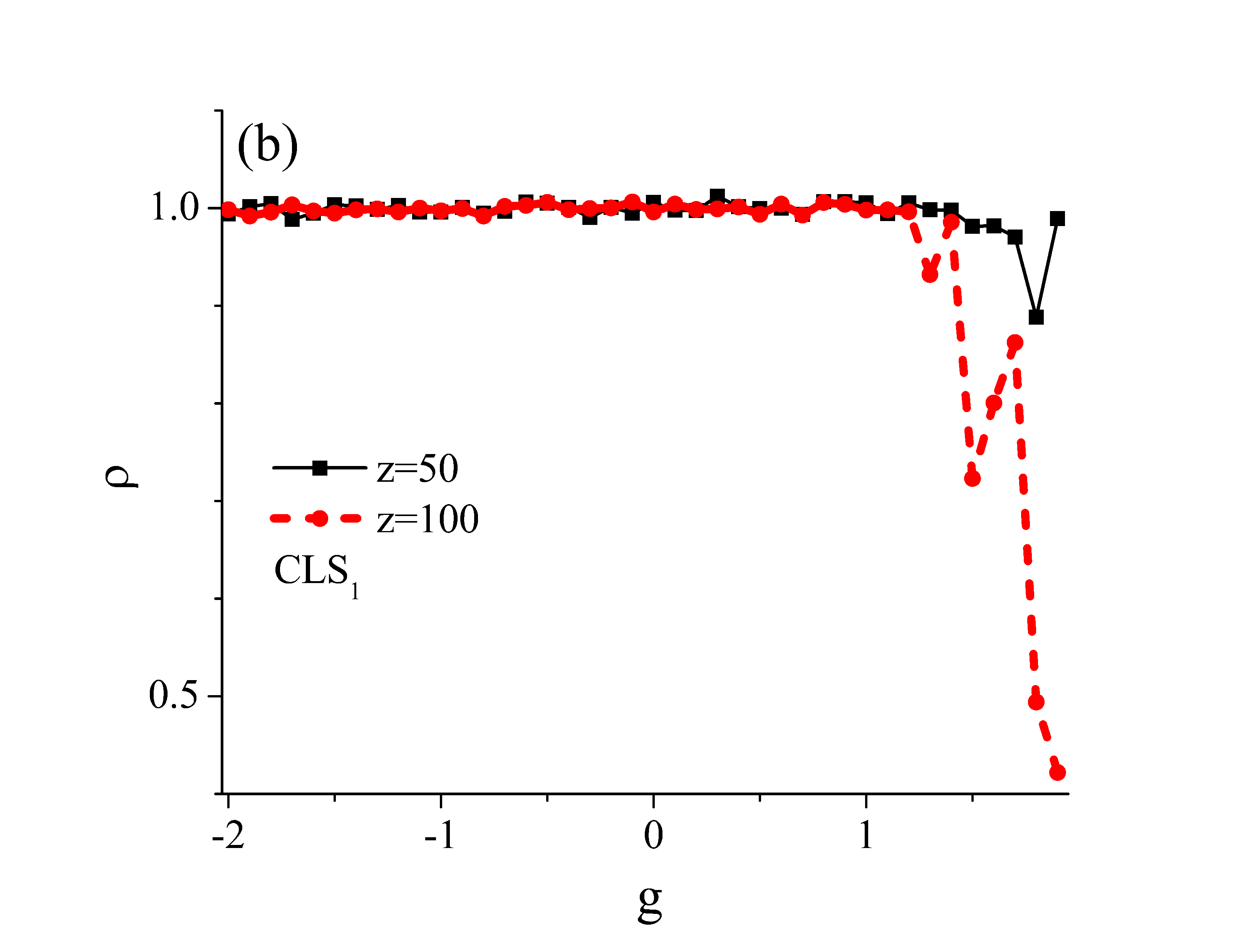}
	\caption{The state overlap $\rho$ for nonlinear $CLS_{0}$ and $CLS_1$ type CLSs, originating from singular zero energy FB at $\phi=0$ and isolated linear FB at $\phi=\pi/3$ (unity energy), respectively. The corresponding flux values are  $\phi=0$ and $\phi=\pi/3$. Black solid lines with squares show the $\rho$ at $z=50$, and red dashed line with cycles at $z=100$.}
	\label{fig5}
\end{figure}

\begin{figure}[ht!]
\centering	\includegraphics[width=7cm]{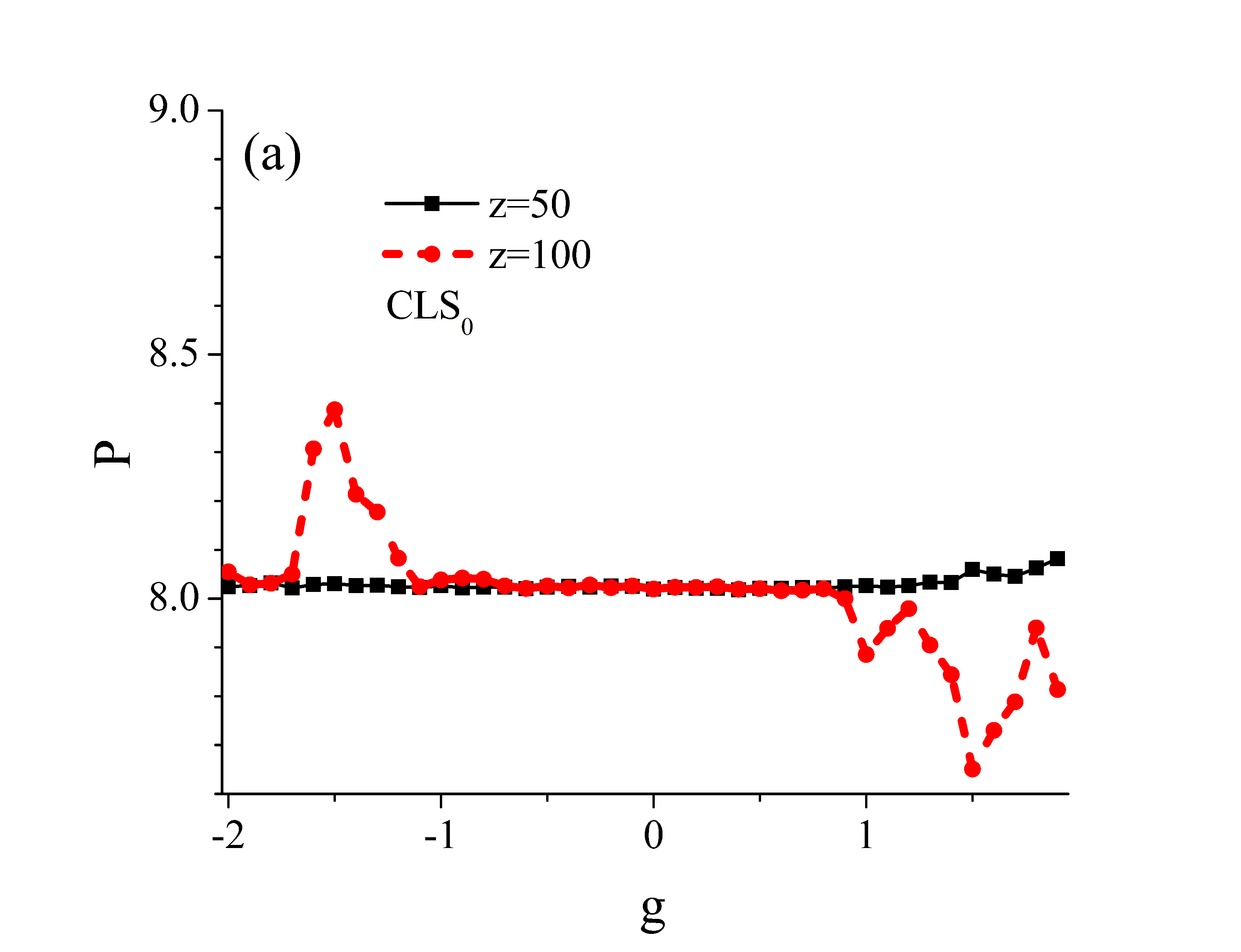}
	\includegraphics[width=7cm]{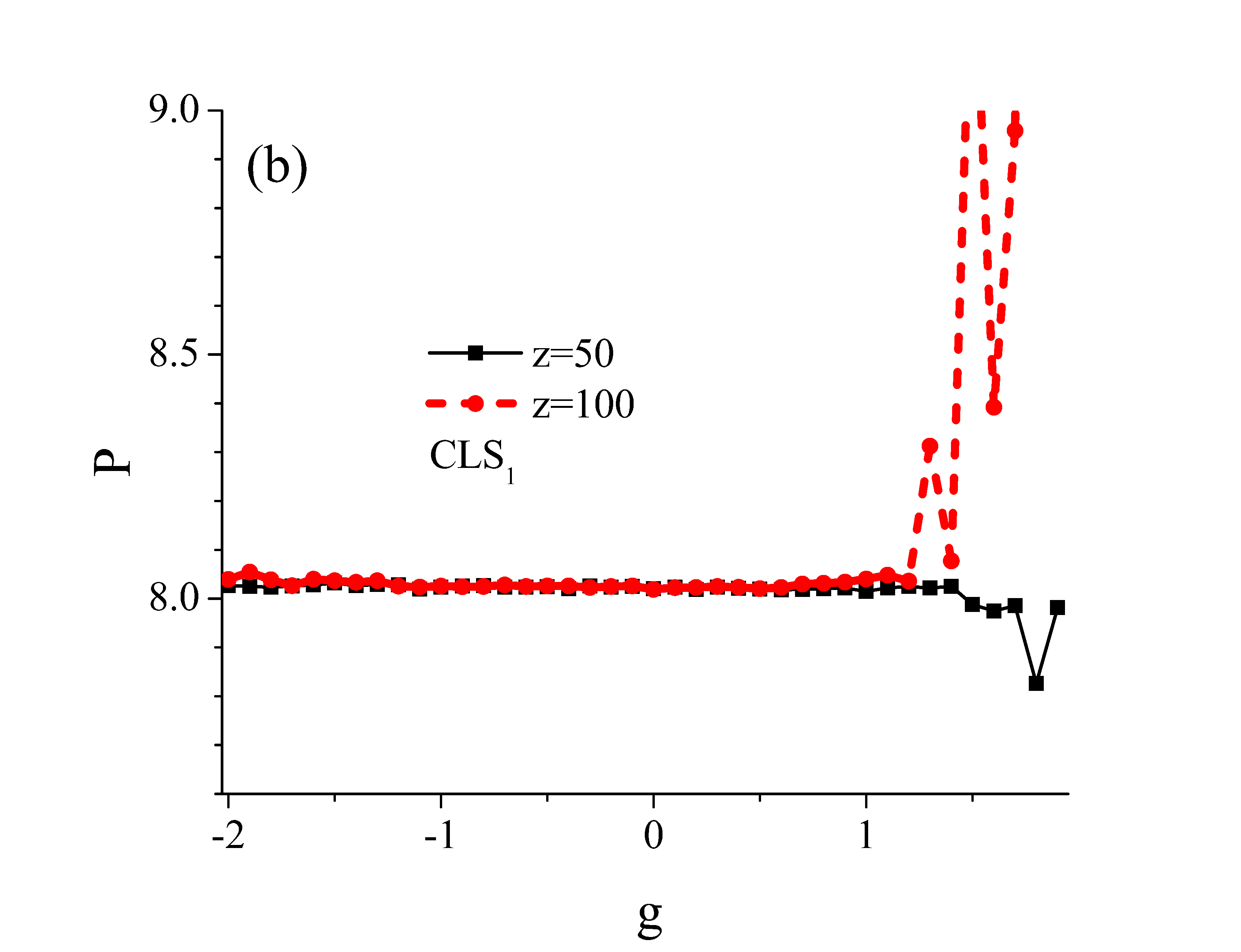}
	\caption{The participation number $P$ for nonlinear $CLS_{0}$ and $CLS_1$ type CLSs, originating from singular zero energy FB at $\phi=0$ and isolated linear FB at $\phi=\pi/3$, respectively. Black solid lines with squares show the $\rho$ and $P$ at $z=50$, and red dashed line with cycles at $z=100$.}
	\label{fig6}
\end{figure}

Dynamical properties of the nonlinear compact localized mode families do not show qualitative peculiarities regarding the type/phase state of the CLS. The clarification of the mode robustness is multiparametrically dependent (parameters: sign and strength of nonlinearity; band width; gap width; flatness). While the strength of nonlinearity is smaller than the gap between the isolated FB and the closest DB (the down/upper band at $\phi=\pi/6$ $(CLS_{-1})$ /$\pi/3$ $(CLS_1)$, respectively) the nonlinear CLS is highly robust, showing the negligible changes in $\rho$ and $P$ in Figs. \ref{fig5} and \ref{fig6}. The changes in both quantities observed for $CLS_1$ in the presence of nonlinearity of both types and $|g|<1$ show stable/weakly unstable propagation. By increasing the defocusing nonlinearity strength ($g>1$) dramatic changes in $\rho$ and $P$ indicate the significant energy radiation over whole lattice during the propagation and destruction of the CLS. On the other hand, in the area of focusing nonlinearity $g<-1$ the $CLS_1$-like pattern is preserved for long propagation distances after balancing its energy with the neighborhood, as can be seen on the state overlap and participation number profiles in Figs. \ref{fig5} and \ref{fig6}. Regarding the nonlinear $CLS_{0}$ equivalent originating from the singular FB, which is the member of robust FB-DB-FB triplet strongly separated from the second DB \cite{nashODL}, the nonlinearity-induced mixing affects the mode overlap for $|g|>1$ in the area of defocusing nonlinearity, as can be confirmed by the profile of the $\rho(g)$ for $CLS_{0}$, Fig.~\ref{fig5}. Regarding $P$, with increasing the propagation length we observed the changes in its value over the whole region $|g|>1$ but with maximal relative discrepancy of the order of $1\%$. The increase ($g<-1$) or decrease  ($g>1$) of $P$ can be associated to the increase or decrease of the number of 'excited' lattice sites, respectively. In the latter case a profile more strongly localized than the initial CLS is formed. 

The reason for different $P$ behaviour during the propagation of the  $CLS_{0}$ and $CLS_{1}$ like nonlinear CLSs can be related to the increased robustness of the structure with phase distribution over $CLS_{1}$ octagon with respect to this over $CLS_{0}$ in the presence of focusing nonlinearity.  On the other hand, in the area $g>1$ more dramatic changes are observed in the case of $CLS_1$ compared to those for $CLS_{0}$. The equivalent conclusions were obtained for the $CLS_{-2}$ and $CLS_{-1}$ compact states. This finding is not fully consistent with the instability growth rate in the early stages of the CLS instability extracted by the LSA, according to which, for CLSs continued from the $CLS_{0}$ and $CLS_{\pm 2}$ (eigenmodes of the FBs caged inside the robust triplet FB-DB-FB) marginal (linear) stability is predicted in the huge area of the negative $g$, while for the CLSs continued from the isolated FBs only in the very close neighborhood of $g=0$. These discrepancies we can associate with the peculiarity of the modes originating from the highly degenerated manifolds (as FBs are). The correct interpretation of the mode behaviour in the areas of neutral/marginal stability reported by the LSA in that cases deserves the higher order perturbation analysis, i.e. the fully nonlinear approach. 

We conclude that in general for experimentally relevant parameters the CLSs radiate negligible energy to the lattice background in the presence of weak nonlinearity and random perturbations. The reason is the caging effect dominance over the weak mixing of the band states driven by nonlinearity-related instability. With increasing nonlinearity strength, the nonlinear mixing between and within bands becomes stronger, which is reflected in the mode dynamics. Note that  here the propagation distance $z\le 100$ corresponds to the experimentally-feasible propagation lengths. 

Finally, we notice that the structured beam nature of the CLS provides the excitation of a wider region of wavevectors of the band which initially hosts the CLS then the plane waves. Developing of instability induced by nonlinearity, extends the excited region of the band and can populate entire band which is simultaneously no more strictly flat, being affected by the nonlinearity too. This opens opportunity to think on utilization of the CLS for scanning the band topological properties in the cases with nearly FBs. In other words, the large area of band will be already populated by the initial state, which potentially allows the Chern number to be measured in a shorter propagation distance then in the case with band excitation driven by the plane wave (the nonlinear Bloch mode)~\cite{topo_MI}.

\subsection{Dynamical properties of compact localized mode's driven by artificial flux modification}

Further we probe the response of the nonlinear compact localized modes to the flux (hopping strength) changes, emulating the externally-driven lattice. The way CLSs adapt to external driving can open a new perspectives regarding the CLSs application in the controlled transport of energy/information in photonic lattices.

\begin{figure}[ht!]
\centering	\includegraphics[width=6cm]{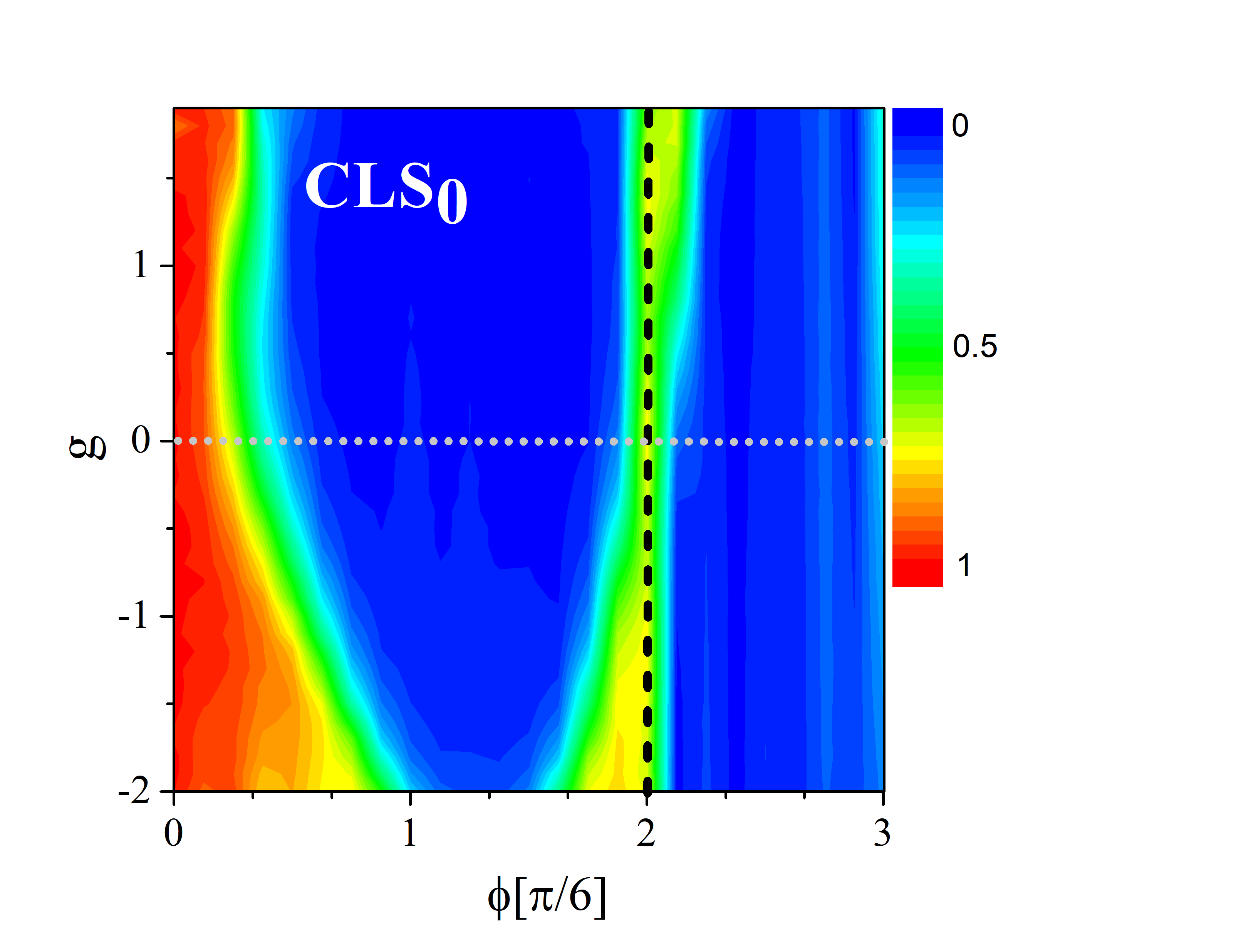}
	\includegraphics[width=6cm]{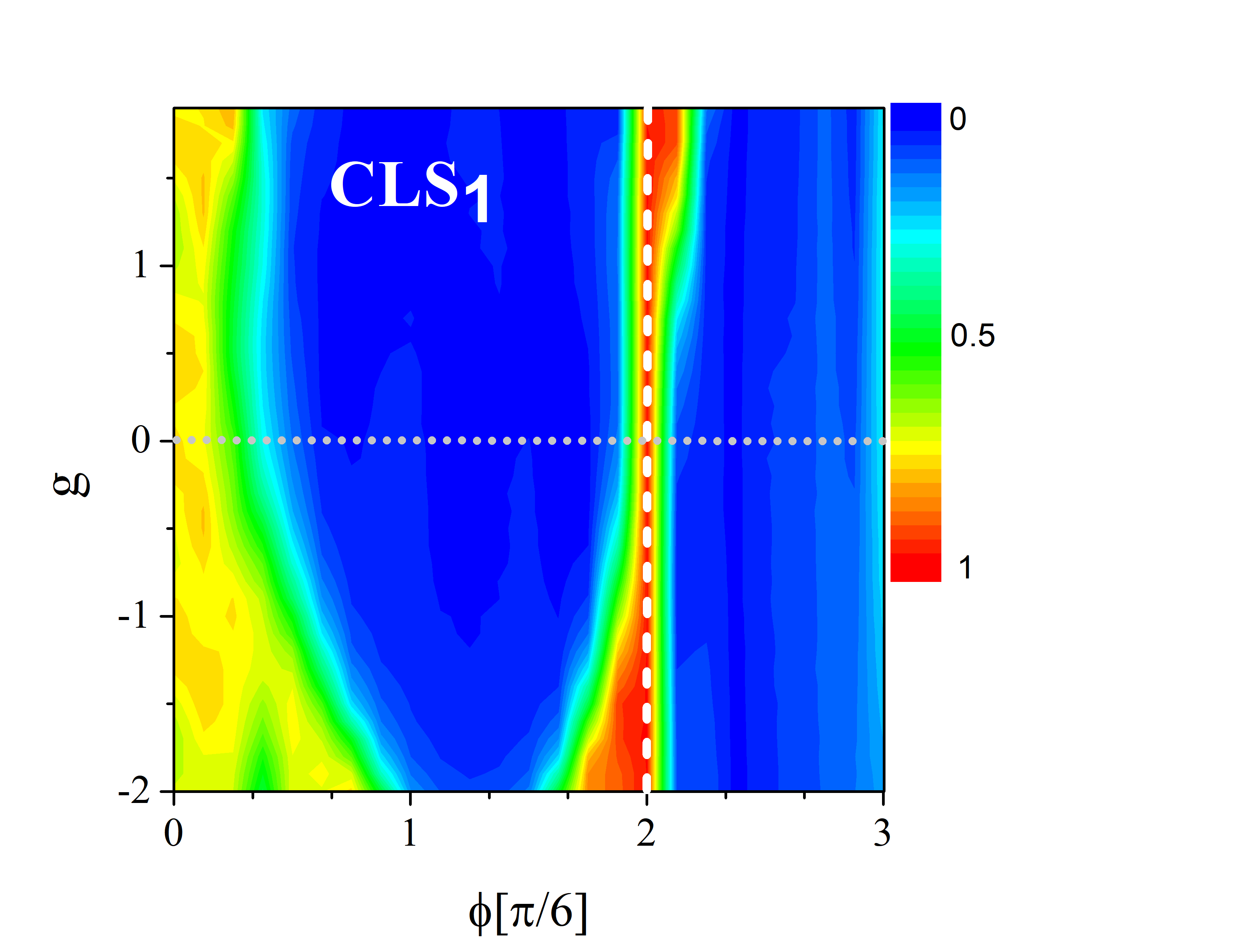}
	\caption{The magnitude of the state overlap $\rho=|<\psi(z=50) | \psi(0) >|^2$ in the parameter space $(\phi,g)$ for $CLS_{0}$ and $CLS_1$ nonlinear compact localized modes. The vertical dashed lines denote the positions $\phi=\pi/3$, while the horizontal dotted line indicates $g=0$ case. Red areas correspond to $\rho=1$, i.e. to stable (weakly unstable) propagation of the NL CLS. }
	\label{fig7}
\end{figure}

\begin{figure}[ht!]
\centering	\includegraphics[width=6cm]{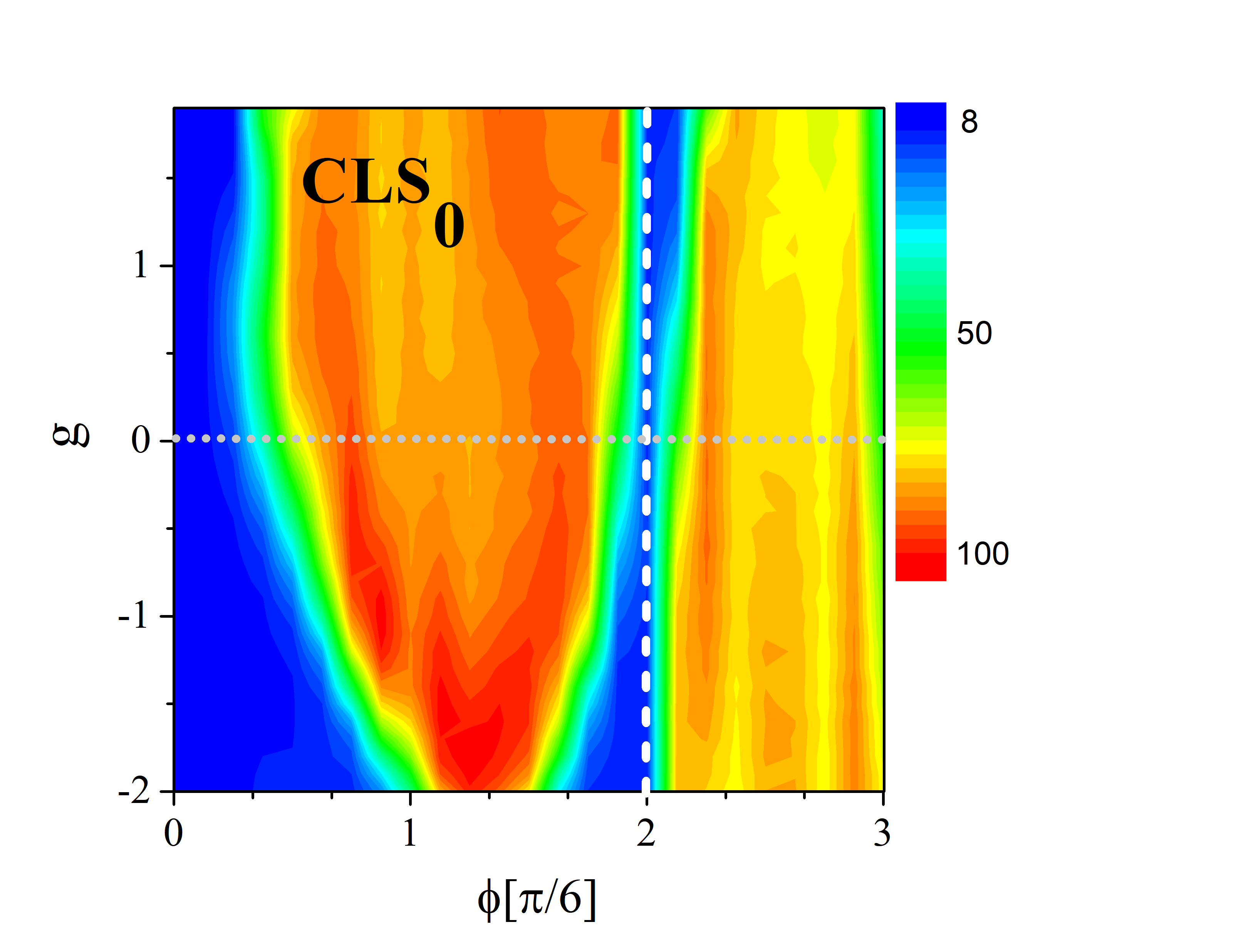}
	\includegraphics[width=6cm]{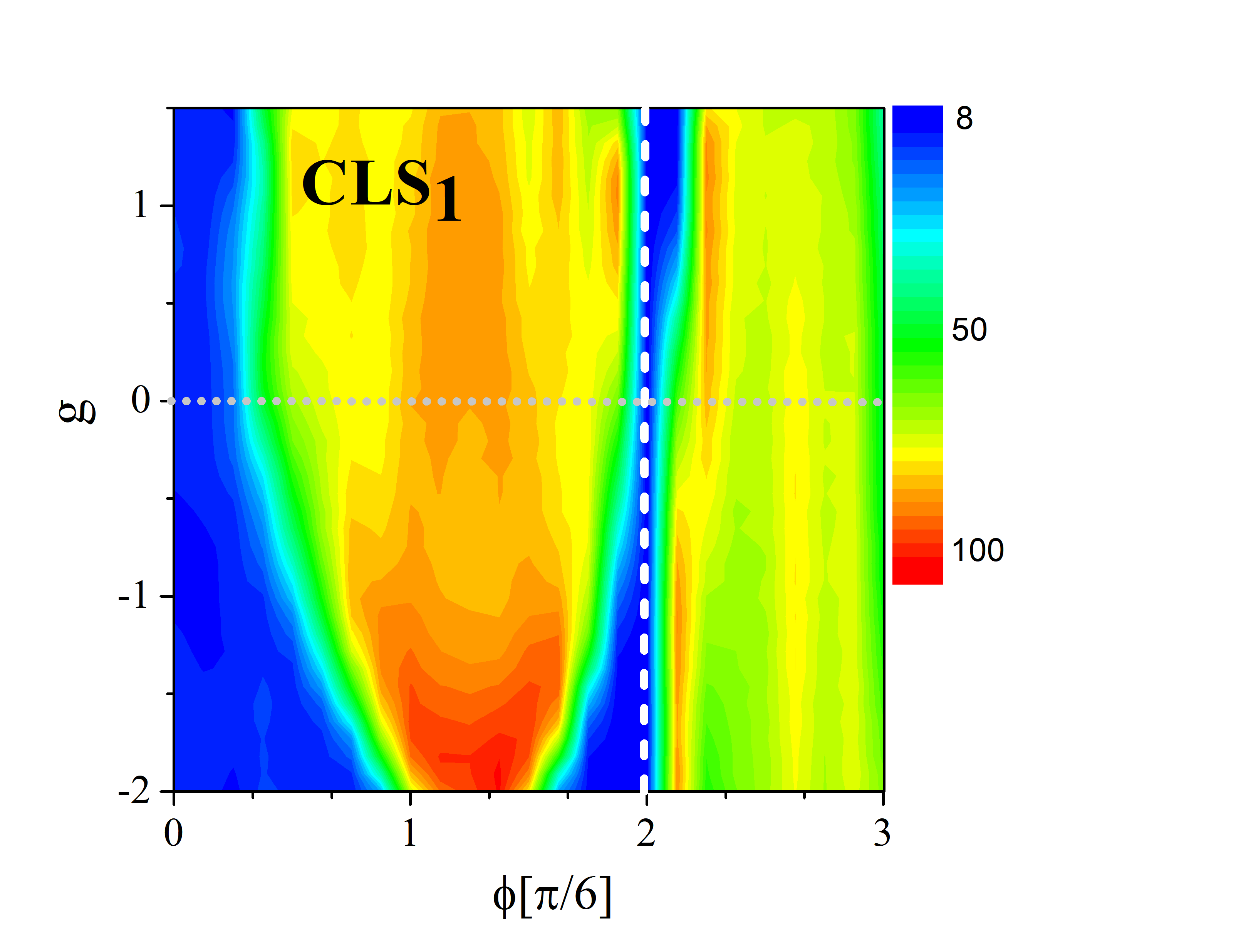}
	\caption{The participation ratio at $z_{end}=50$ in the parameter space $(\phi,g)$ is shown for $CLS_{0}$ and $CLS_1$. Two vertical dashed lines denote the positions $\phi=\pi/3$, while the horizontal dotted line is guide for eyes indicating the absence of nonlinearity. Dark blue areas correspond to the weakly unstable (stable) CLS propagation.
	}
	\label{fig8}
\end{figure}

In general, the CLSs are highly sensitive to any change in flux as shown in Figs. \ref{fig7} and \ref{fig8}, which present the field overlap and participation ratio at $z=50$ over the $(\phi,g)$ parameter space for the nonlinear $CLS_{0}$ originating from $\beta=0, \phi=0$ FB, and the nonlinear $CLS_1$ equivalent originating from $\beta=1, \phi=\pi/3$ FB. The change of flux detunes the conditions for perfect caging which is sensed via dephasing of field components within the CLSs' assigned octagon, and radiation of mode energy to the rest of lattice. The maximum field overlap and $P=8$ for $CLS_{0}$ is at $\phi=0$ and at $\phi=\pi/3$ for $CLS_1$. In general, the rate of energy exchange will be the highest in the neighborhood of critical points for all CLSs, as confirmed by the $\rho\rightarrow 0$ and $P\rightarrow 100$ at $\phi=\pi/4$ (blue and red areas, respectively) in Figs. \ref{fig7} and \ref{fig8} at selected fixed value of $g$. 

\begin{figure}[ht!]
\centering	\includegraphics[width=15cm]{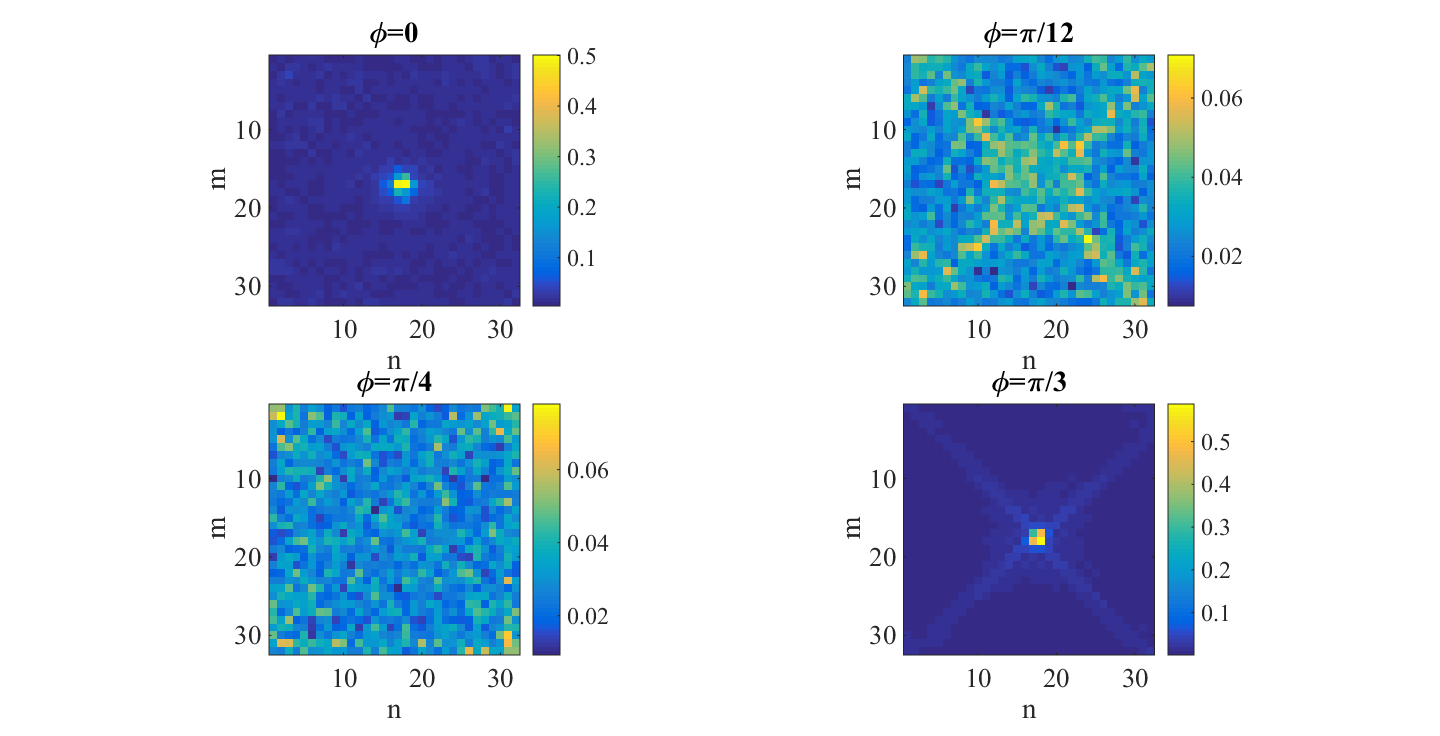}
	\caption{Total mode intensity distribution in real space at the end of simulation for compactons of type $CLS_1$ in the presence of different flux values.
		The nonlinearity parameter is $g=1$. }
	\label{fig9}
\end{figure}

Interestingly, by changing flux we observed recovery of the robustness of the CLS in the areas of the $\phi_{FB}$, i.e. $\phi=\pi/3$ and $\phi=0$ for $CLS_{0}$ and $CLS_1$ nonlinear compact localized mode equivalents, respectively, in the presence of nonlinearity of weak (and moderate) strength (cases plotted here).  In other words, the propagating nonlinear CLS of the $CLS_{0}$ type, shown in Fig. \ref{compacton}, is 'recaptured' to the initial octagon ring at $\phi=\pi/3$. The corresponding magnitude of the field overlapping is $\rho(z=50)\approx 0.7$. The value of the participation number $P\approx 12$, while the initial one was $P=8$, indicates the limited spreading of the CLS through the propagation length $z=50$.
The equivalent situation occurs with the nonlinear CLS of the type $CLS_{1}$ (the eigenmode for FB at $\phi=\pi/3$), which recovers its integrity after switching the flux on the value $\phi=0$, where the $CLS_{0}$ and $CLS_{-2}$ compactons are supported in the linear case. The $\rho\approx 0.6$ ($0.7$) shows the efficiency of the energy recovering at the starting CLS's octagon position. $P$ is around $1.2$ times higher than the initial value. The robustness of the CLSs on the flux values can be also seen in Fig. \ref{fig9}, which shows the total mode intensity at certain $z$ for a few selected values of the flux. 

In summary, tailoring the flux and nonlinearity the flux-driven ODL CLSs could be considered as a promising tool for controlling the energy transport through the lattice and probing the band structure of the underlying lattice. 

\section{Concluding remarks}

The flux-dressed ODL provides a toy model for generating the topological phase transitions, including between singular and non-singular perfect FBs, and nearly flat topologically nontrivial bands. In the root of these diversity is the Aharonov-Bohm effect as a consequence of the flux tuned hopping inside the lattice. Here we are focused on testing the robustness of compact localized modes originating from different types of perfect flatbands.  This opens a gate towards utilization of the compact localized modes for scanning the properties of flux-driven lattices. In general, we demonstrated the ability of CLS to sense the environmental changes by redistributing the intra-site phases while keeping the compactness. Therefore, by tuning the local gauge flux and site hoppings, the CLS can be taken as a probe of the topological and transport properties of the photonic lattices.  

The instability excited due to the inherent nonlinear response of the photonic media to the laser light propagation provokes the dynamical response of the compact localized modes to the exchanged environment. In the region of weak nonlinearity the CLS families extended either from the linear compacton families of non-singular or singular FBs shown to be highly robust structures. In addition, we showed the nearly stable/weakly instable dynamical propagation of the nonlinear CLSs in certain flux managed configurations which can include topologically nontrivial cases far from the transition points under properly tuned system parameters. In other words, by tailoring the gauge field flux the nonlinear compact localized modes can be used to explore the topology of photonic lattices and to manage diverse optical functions. The recent experiments with the laser inscribed waveguides and ring resonators offer them as suitable platforms for the CLS manipulation.

\section*{Acknowledgements}

This research was supported by the Ministry of Education, Science and Technological Development of the Republic of Serbia  (grant number 451-03-9/2021-14/ 200017) and the National Research Foundation, Prime Ministers Office, Singapore, the Ministry of Education, Singapore under the Research Centres of Excellence programme, the Polisimulator project co-financed by Greece and the EU Regional Development Fund and Institute for Basic Science in Korea (IBS-R024-D1).

\end{document}